\documentclass[conference]{IEEEtran}

\ifCLASSINFOpdf
  \usepackage[pdftex]{graphicx}
\else
\fi

\usepackage{dirtree}
\usepackage{listings}
\usepackage{multirow}
\usepackage{subcaption}
\usepackage{siunitx}
\begin{document}

\title{Dataset for anomalies detection in 3D printing}

\author{
\IEEEauthorblockN{
Joanna Sendorek, Tomasz Szydlo, Mateusz Windak, Robert Brzoza-Woch}
\IEEEauthorblockA{AGH University of Science and Technology,\\
Department of Computer Science, Krakow, Poland\\
Email: send.joanna@gmail.com, tszydlo@agh.edu.pl}
}

\maketitle

\begin{abstract}
Nowadays, Internet of Things plays a significant role in many domains. Especially, Industry 4.0 is making a great usage of concepts like smart sensors and big data analysis. IoT devices are commonly used to monitor industry machines and detect anomalies in their work. In this paper we present and describe a set of data streams coming from working 3D printer. Among others, it contains accelerometer data of printer head, intrusion power and temperatures of the printer elements. In order to gain data we lead to several printing malfunctions applied to the 3D model. Resulting dataset can therefore be used for anomalies detection research. 
\end{abstract}

\IEEEpeerreviewmaketitle

\section{Introduction}

This paper presents the data that we have gathered from the 3D printer during the printing process. Among all, data samples include temperature of working elements of the printer, intrusion force and the acceleration of printing head. The data has been gathered using two types of sources - custom-made measurement devices and the internal software of the printer.


In order to enable for the dataset to serve as an example of anomalies detection for intelligent Industry 4.0 systems, we provoked several types of failures during printing process. All of the files are placed in the repository\footnote{https://github.com/joanna-/3D-Printing-Data} and can be used under the \texttt{Creative Commons Attribution 4.0 International} license.

The rest of the paper is organised as follows. Section \ref{related} presents related work. In section \ref{printer} we describe the characteristics of printer machine used for gathering data samples. In \ref{data} we characterize each type of data source while  printing failures that we created are presented in \ref{types}. Section \ref{org} contain specification of data organisation while section \ref{analysis} contains exemplary data analysis. The last section sums up the paper.

\section{Related Work}\label{related}
As an interest in 3D printing increases in various applications,  anomaly detection systems are gaining in importance. We can basically distinguish two types of systems used to monitor the work of 3D printers and received printouts. These systems are based on image analysis (e.g. \cite{8997053,refId0}) and based on the analysis of data from inertial sensors (e.g. \cite{7222632, 8461266}). 

Proper preparation of the 3D printer and retrofitting it with sensors requires some time and equipment expenditure. In the article we present a set of test data that were obtained using devices built using the FogDevices platform. The presented set of test data can be used to develop new algorithms for detecting anomalies in the work of 3D printers and, what is important, to compare them.

\section{3D Printer characteristics}\label{printer}


The 3D printer utilized for collecting its operation data was Monkeyfab Spire manufactured by Monkeyfab\footnote{http://www.monkeyfab.com} - its basic properties are listed in Table \ref{tab:printerparams}. It is a \emph{delta printer} in which the printing head is mounted on magnetic ball joints. The Monkeyfab Spire uses the \emph{RepRapFirmware} and is controlled over the network via \emph{Duet Web Control}\footnote{https://duet3d.dozuki.com/Wiki/Duet\_Web\_Control\_Manual (access for 20.11.2019)} interface.

\begin{table}[!htbp]
\centering
\caption{Basic parameters of the utilized 3D printer \\ according to manufacturer's specifications.}
\label{tab:printerparams}
\begin{tabular}{|l|l|}
\hline
Maximum printed object dimensions & \begin{tabular}[c]{@{}l@{}}\SI{150}{\milli\metre} diameter\\ \SI{165}{\milli\metre} height\end{tabular} \\ \hline
Default nozzle diameter           & \SI{0.4}{\milli\metre}  \\ \hline
Minimum layer height              & \SI{0.05}{\milli\metre}  \\ \hline
Filament diameter                 & \SI{1.75}{\milli\metre} \\ \hline               Maximum hotend temperature        & \SI{262}{\celsius} \\ \hline                 Maximum platform temperature      & \SI{120}{\celsius} \\ \hline                                                   
\end{tabular}
\end{table}

\section{Data sources characteristics}
The sensor data comes from two sources - (i) internal electronics that control the operation of the printer and from (ii) additionally mounted sensors. They are described in more details in the next subsections.

\label{data}
\subsection{Duet Web Control}

\texttt{Duet Web Control Interface} is the user interface (UI), accessible via web browser, that allows to monitor and change printer state. Among others, it includes such features as: emergency stop, monitoring temperatures of printer parts, changing filament and selecting 3D models to print.
The aforementioned informations are also exposed via API in json format which can be accessed at \textit{[printer server address]/rr\_status?type=X}, where X is the category of data format. In the created datasets, we have used two categories -  \textbf{1} and \textbf{3} alternatively. The core data provided by the API is the same for these categories, but they differ in some extra information. For example, aforementioned third version includes currently printed layer of the 3D model.

\subsection{Data acquisition hardware}
The printer has been equipped with additional custom sensors developed as part of the FogDevices\footnote{http://fogdevices.agh.edu.pl} research project. Data from them was collected using a device assembled using modular hardware components.

\begin{figure}[!htbp]
  \centering
  \includegraphics[width=0.9\columnwidth]{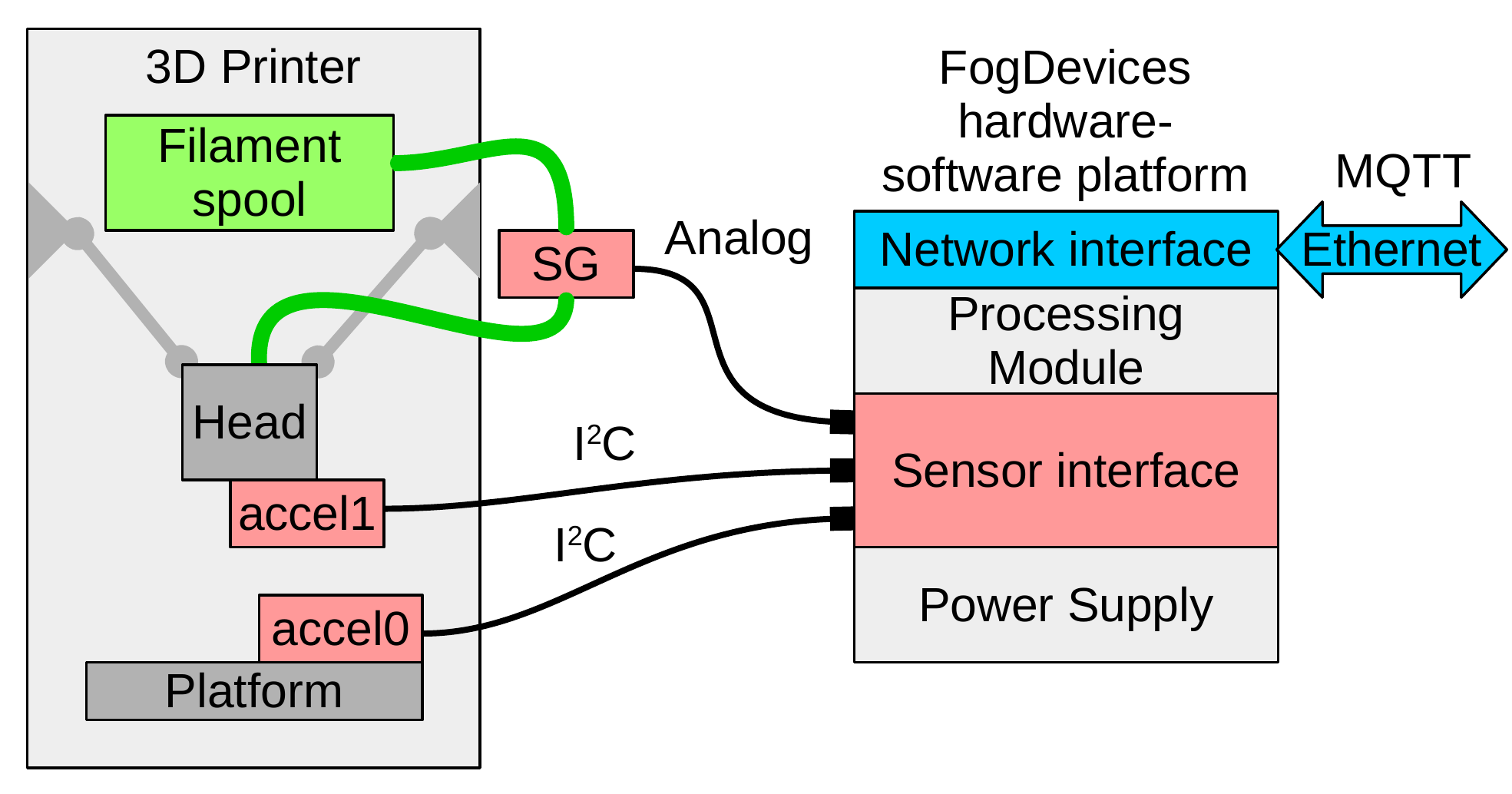}
  \caption{The utilized data acquisition system.}
  \label{fig:test_setup}
\end{figure}

The printer has been equipped with two inertial measurement unit (IMU) sensors LSM9DS1 that can measure acceleration, angular rate and magnetic field in 3 axis but only linear acceleration was used in this case. First of the sensors, called \emph{accel0} is attached to the printing platform and \emph{accel1} is on the print head. Both of the sensors use the I\textsuperscript{2}C digital interface and are connected to the FogDevices hardware platform.

The method of measuring the filament feeding force is based on indirect measurement of the force acting on the Bowden tube during printer operation. This was possible due to the fact that the extruder is located on the body of the printer, not at the print head.
Therefore, a force sensor \emph{SG} based on a strain gauge was developed. Its operation is based on Wheatstone bridge and it produces small voltage output. The voltage is amplified in FogDevices sensor interface module with INA128 instrumentation amplifiers and then measured using an analog-to-digital converter (ADC) with 12-bit resolution.


Block diagram of the hardware is presented in figure~\ref{fig:test_setup}. The FogDevices hardware platform has been utilized to collect data from three sensors: \emph{SG}, \emph{accel0}, and \emph{accel1}.



Data collected by the device was being sent through the MQTT protocol over the Ethernet interface. The data were then saved by a data logger running on a PC computer. The acquisition system collects and processes 200 samples per second.

Additional sensors and devices are provided by the FogDevices platform. The video showing printing process is available online\footnote{https://youtu.be/SFBInVsVDgk}.
\section{Type of prints} \label{types}

\begin{figure*}[!htbp]
    \begin{center}
    \begin{subfigure}[t]{0.4\textwidth}
        \includegraphics[width=\textwidth]{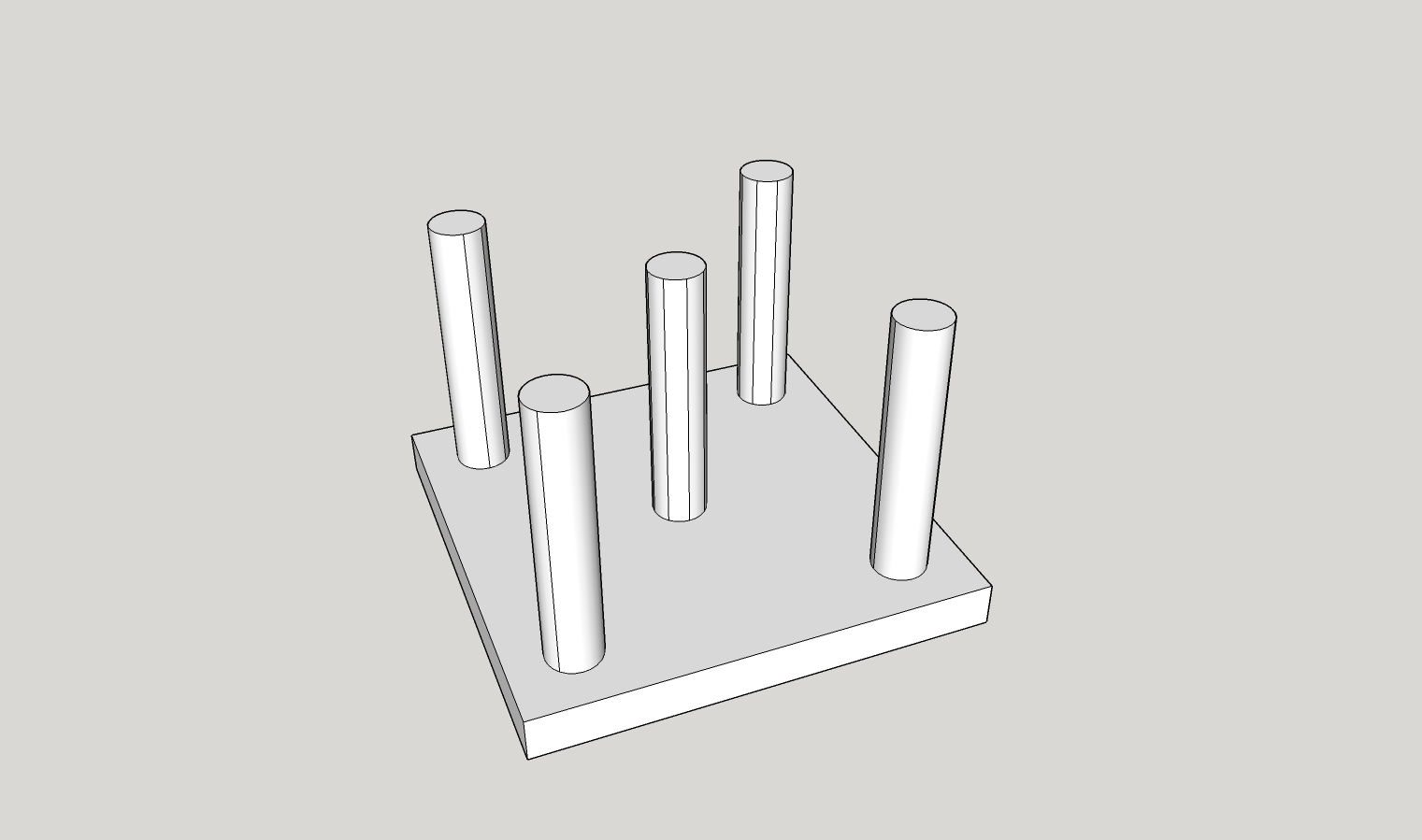}
        \caption{towers with the base - schema}
        \label{base}
    \end{subfigure}
    \begin{subfigure}[t]{0.4\textwidth}
        \includegraphics[width=\textwidth]{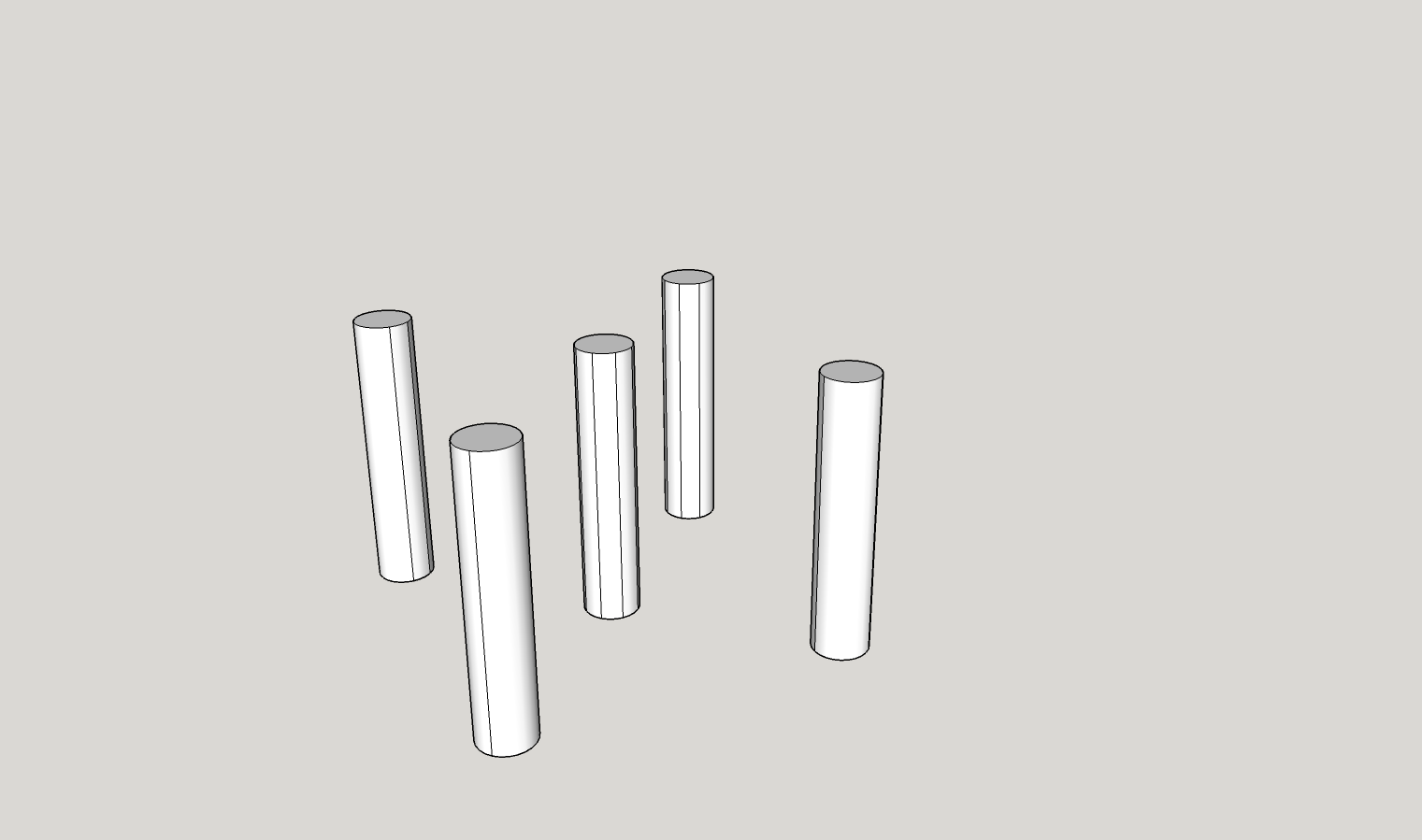}
        \caption{towers without base - schema}
        \label{nobase}
    \end{subfigure}
    \end{center}
    \begin{center}
    \begin{subfigure}[t]{0.4\textwidth}
        \includegraphics[width=\textwidth]{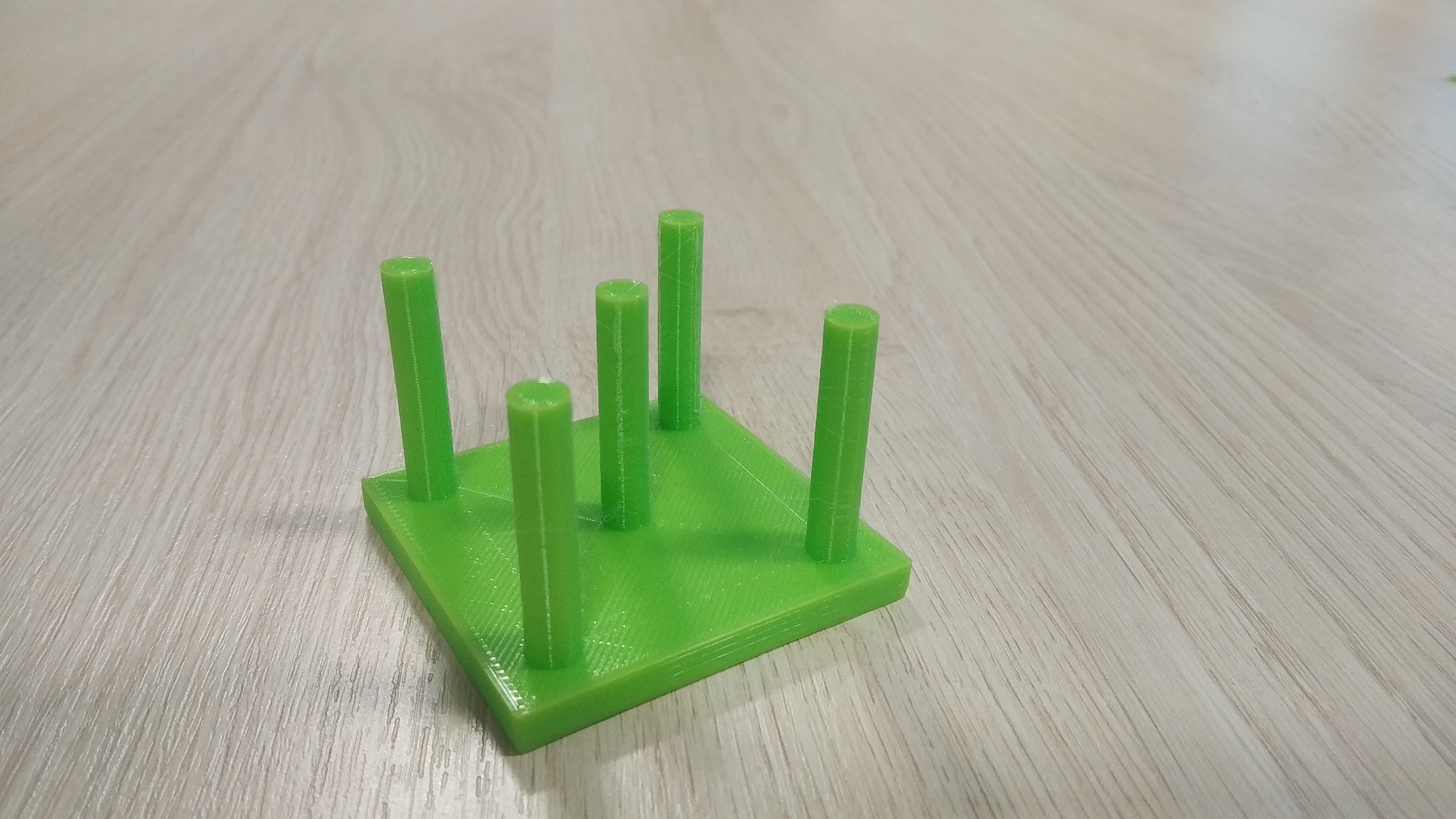}
        \caption{towers with the base - photo}
        \label{basephoto}
    \end{subfigure}
    \begin{subfigure}[t]{0.4\textwidth}
        \includegraphics[width=\textwidth]{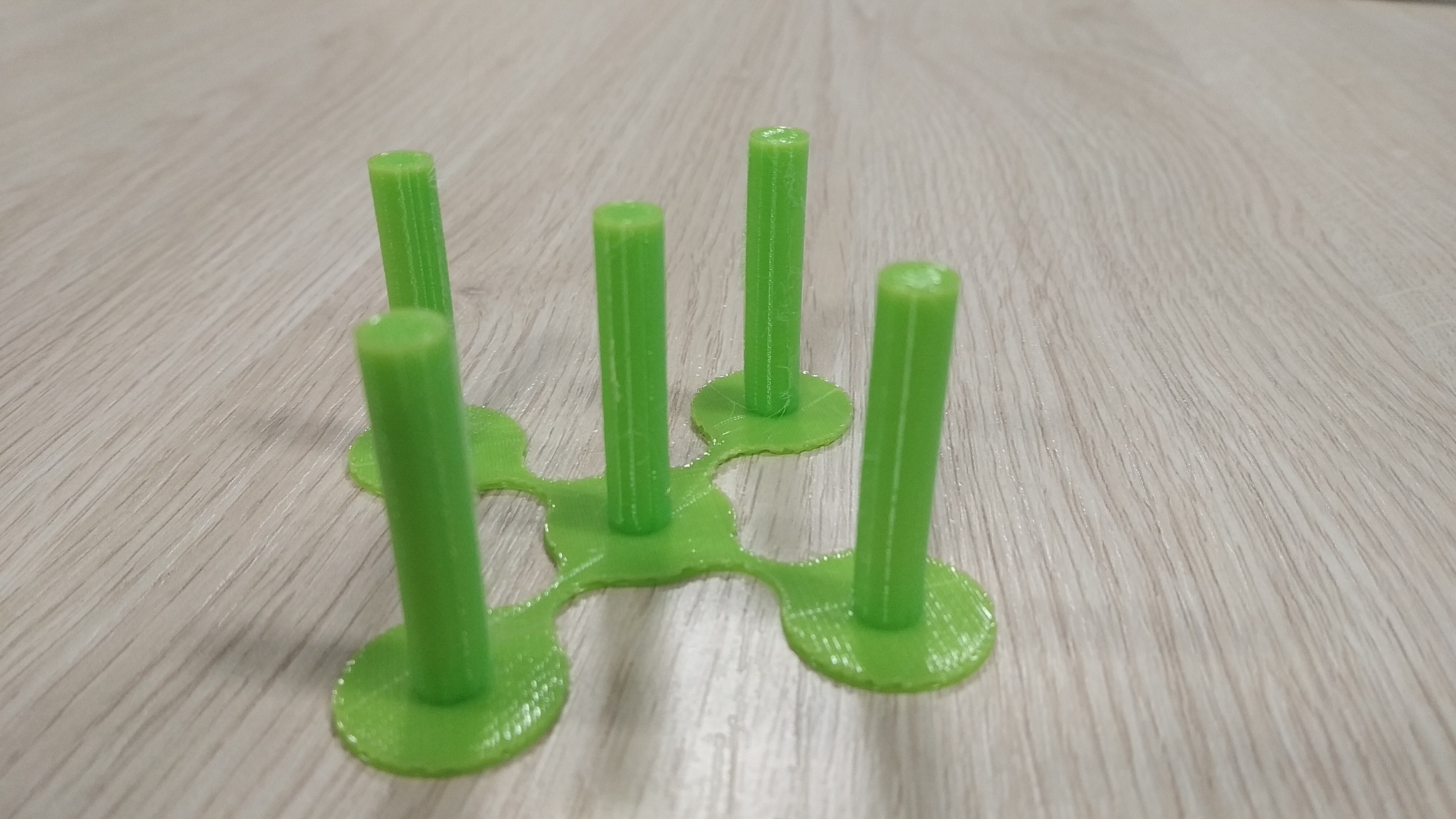}
        \caption{towers without base - photo}
        \label{nobasephoto}
    \end{subfigure}
    \end{center}
\caption{Printing schemas}
\end{figure*}

We have used two variants of the same five towers print in order to collect data. In the variant (a), presented in figures \ref{base} and \ref{basephoto}, towers have printed base that is integral part of the print and in variant (b) presented in figures \ref{nobase} and \ref{nobasephoto}, towers do not have a base - they are placed only on the raft.

\begin{figure*}[!htbp]
    \begin{center}
    \begin{subfigure}[t]{0.32\textwidth}
        \includegraphics[width=\textwidth]{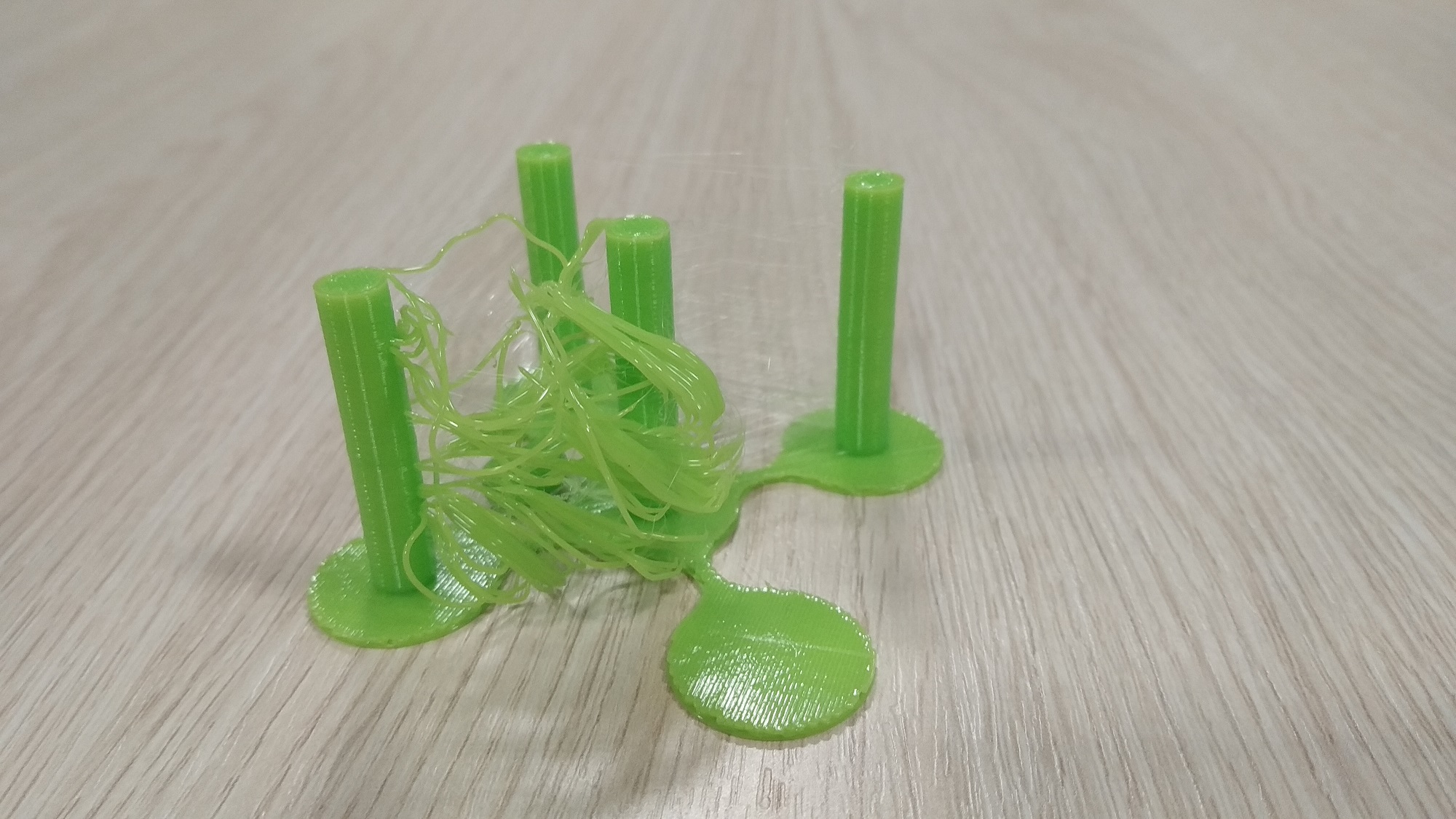}
        \caption{removal of the part of the print}
        \label{removal}
    \end{subfigure}
    \hfill
    \begin{subfigure}[t]{0.32\textwidth}
        \includegraphics[width=\textwidth]{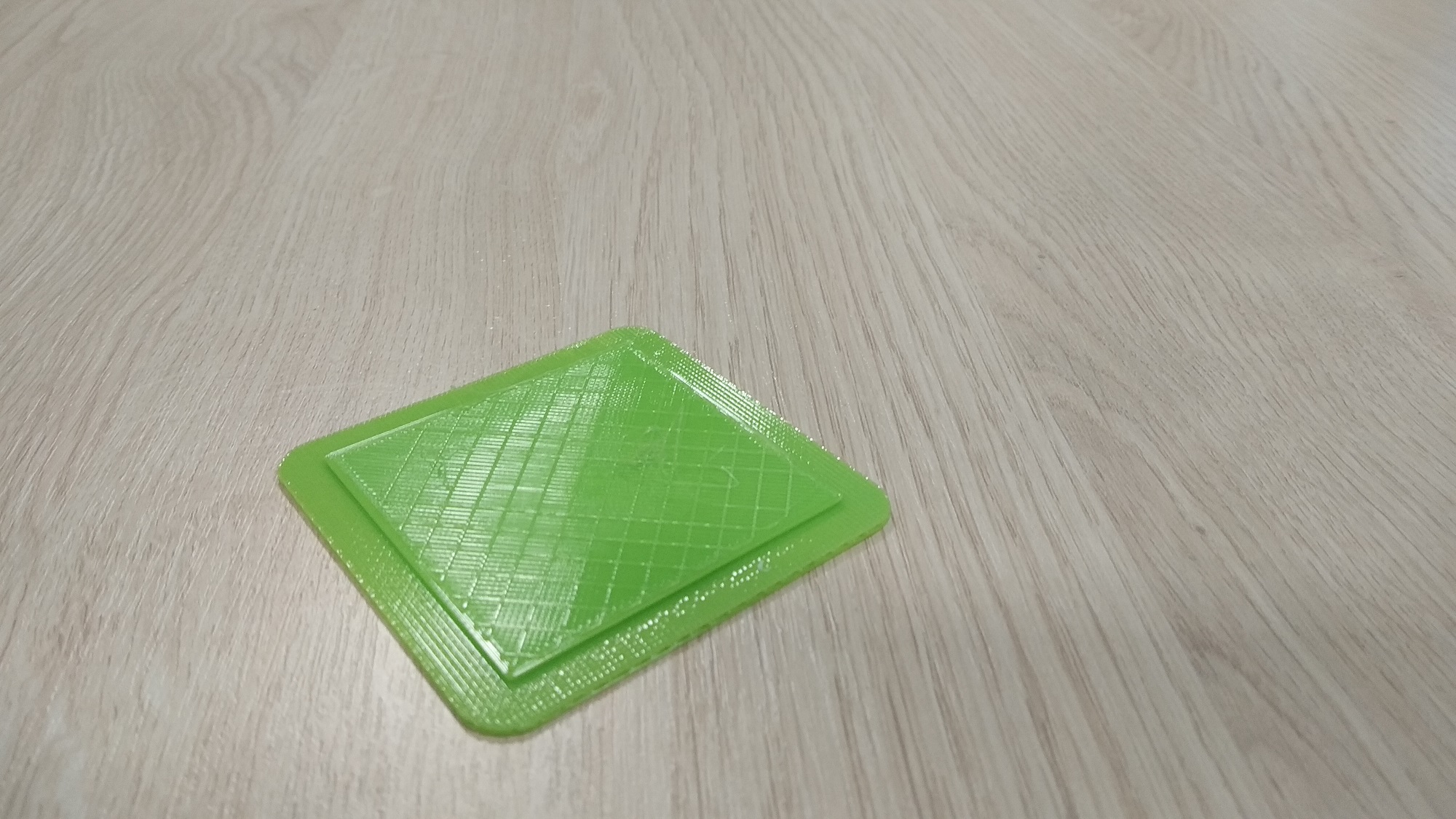}
        \caption{plastic finish}
        \label{plastick}
    \end{subfigure}
    \hfill
    \begin{subfigure}[t]{0.32\textwidth}
        \includegraphics[width=\textwidth]{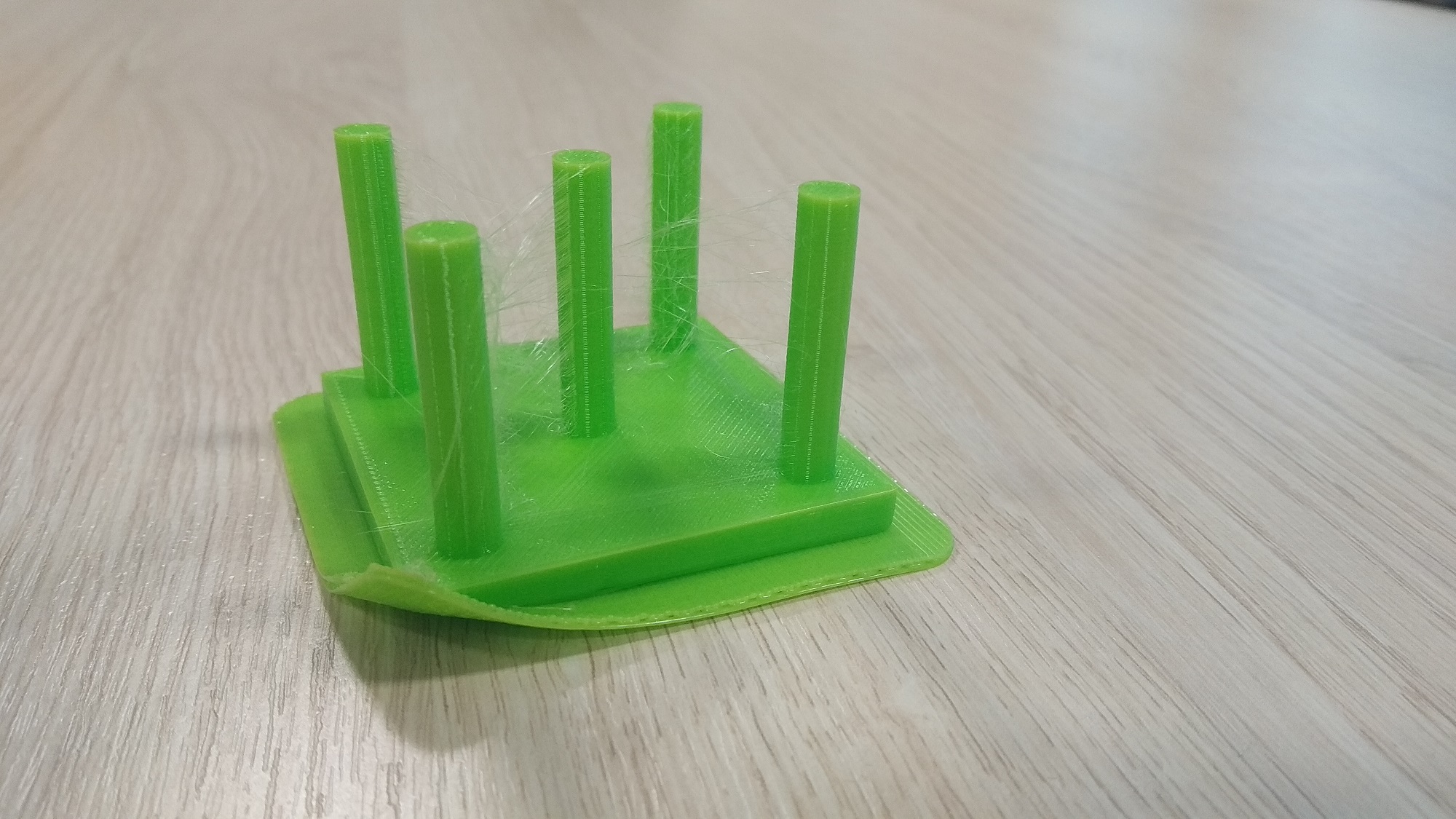}
        \caption{print unsticking}
        \label{unstick}
    \end{subfigure}
    \end{center}
    \begin{center}
    \begin{subfigure}[t]{0.32\textwidth}
        \includegraphics[width=\textwidth]{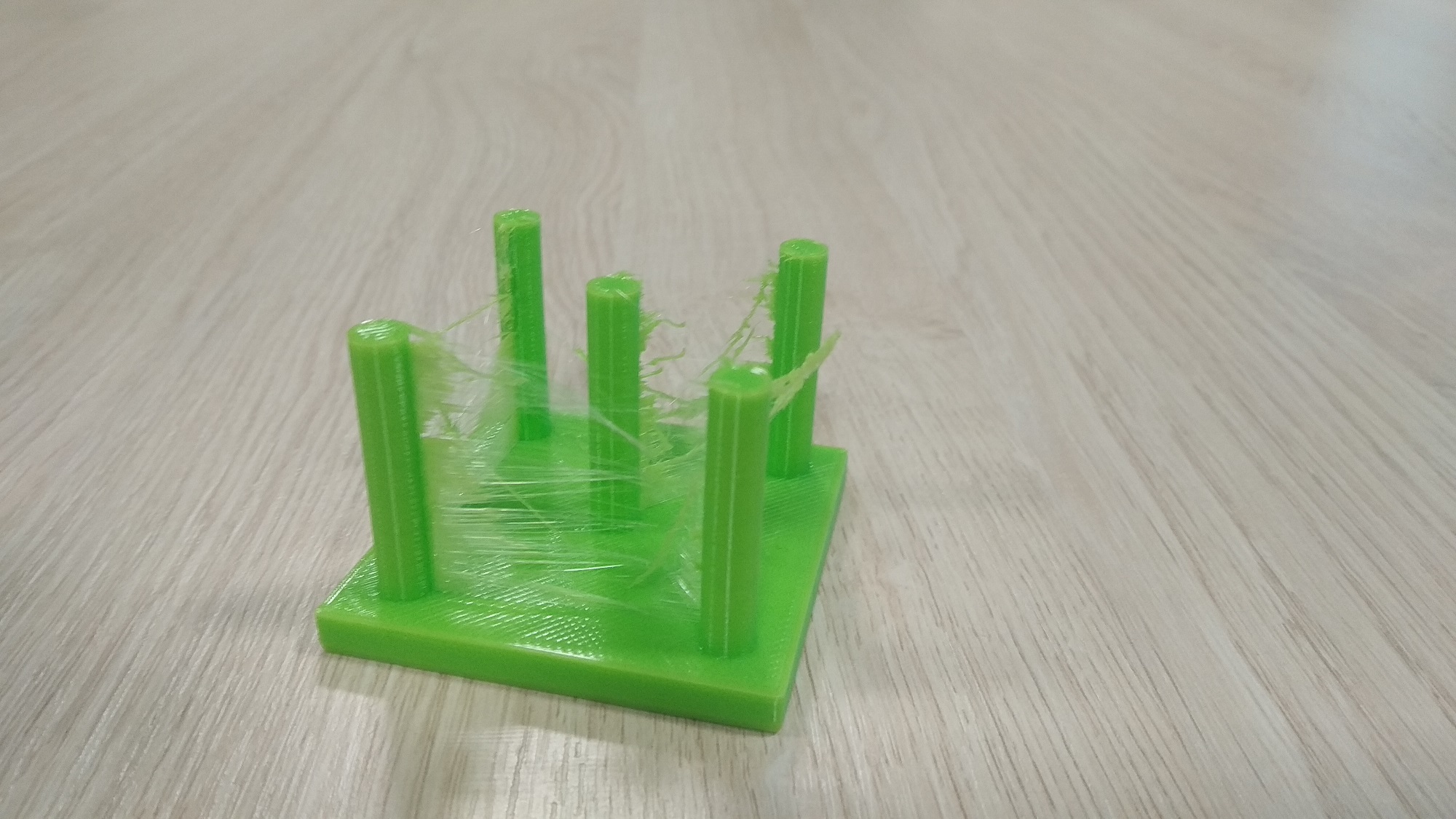}
        \caption{0.5 retraction}
        \label{retraction}
    \end{subfigure}
     \hfill
     \begin{subfigure}[t]{0.32\textwidth}
        \includegraphics[width=\textwidth]{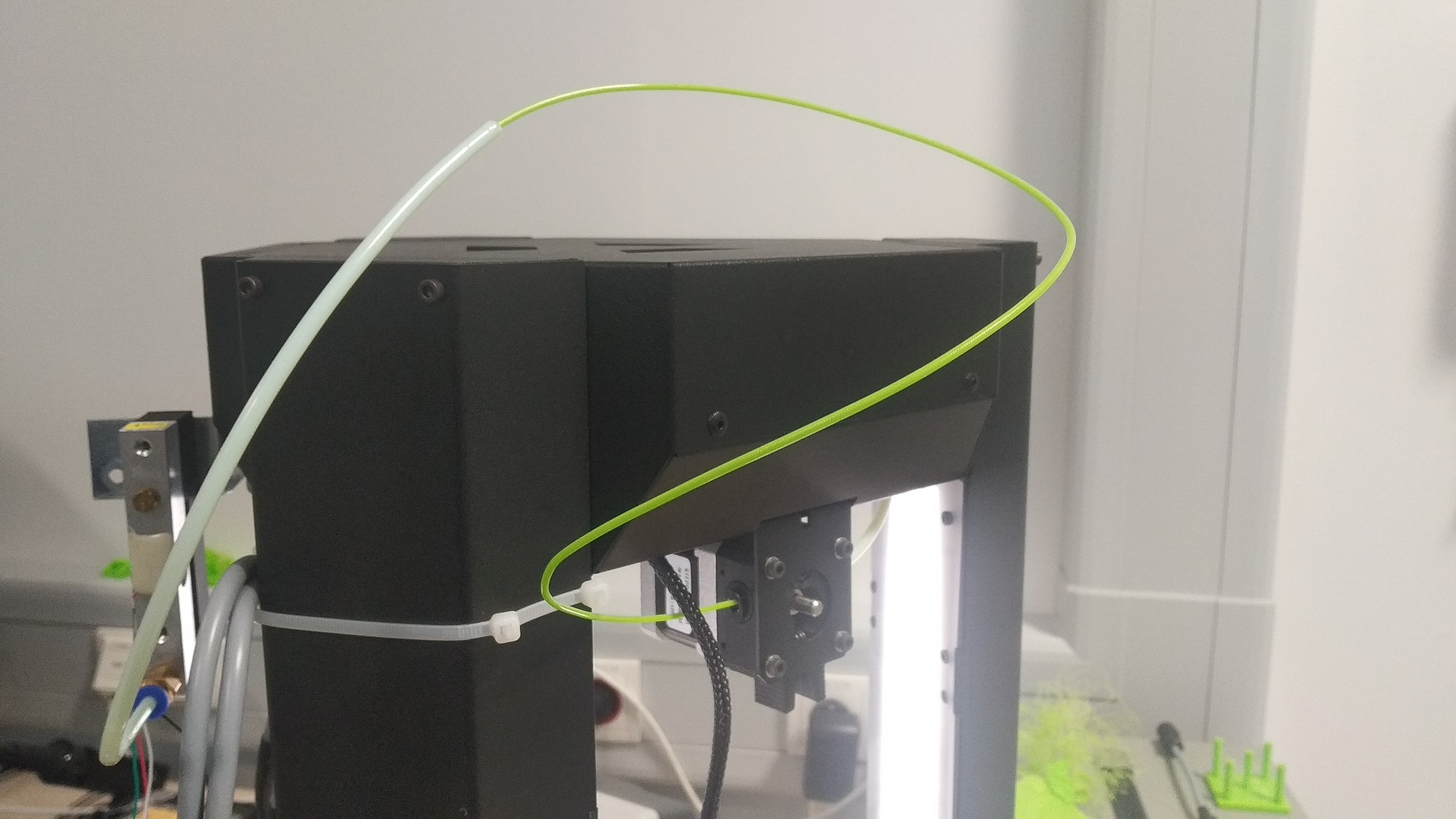}
        \caption{Bowden tube fallout}
        \label{bowden}
    \end{subfigure}
    \hfill
    \begin{subfigure}[t]{0.32\textwidth}
        \includegraphics[width=\textwidth]{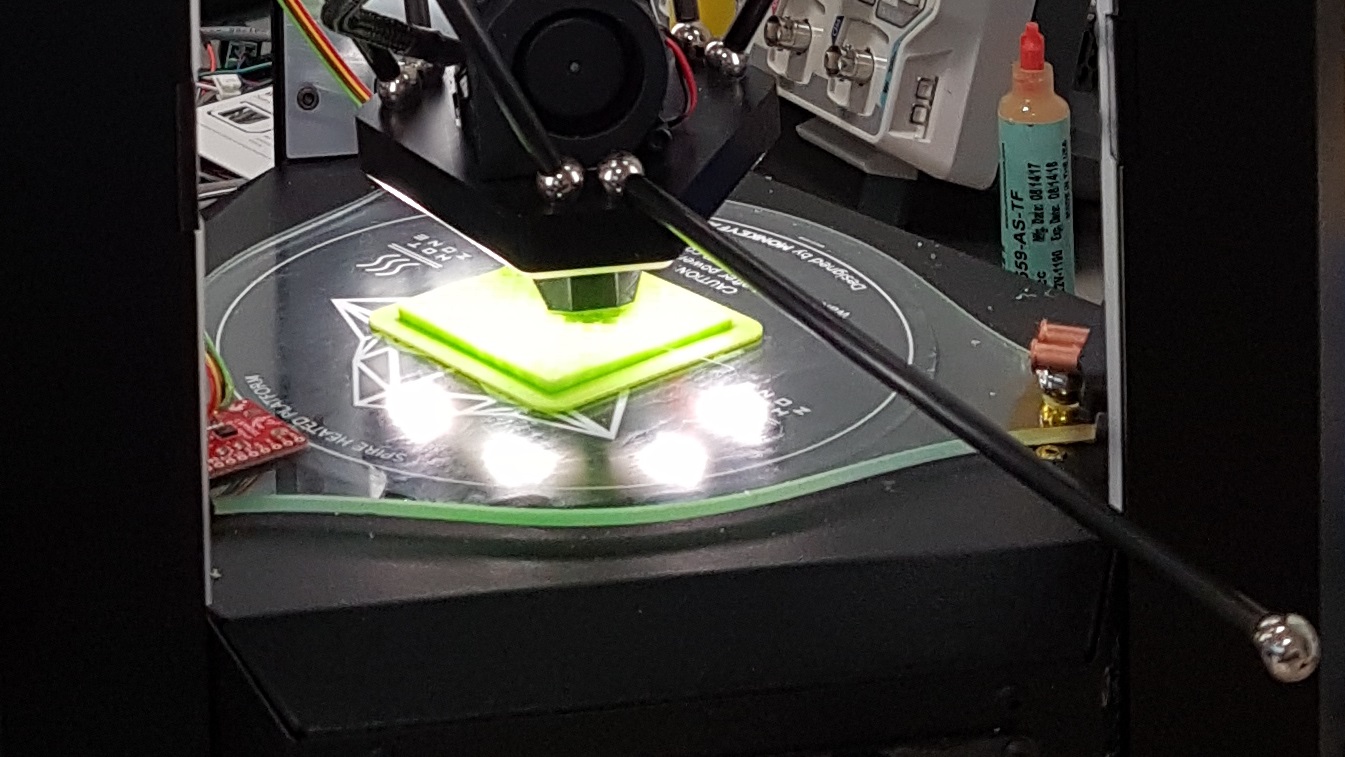}
        \caption{arm failure}
        \label{arm}
    \end{subfigure}
    \end{center}
\caption{Various malfunctions of the print}
\label{anomalies}
\end{figure*}

For both variants, we have collected data from the undisturbed, properly made print. Apart from that, we provoked six printing anomalies presented in figure \ref{anomalies}:

\begin{enumerate}
    \item \textbf{variant (a)}:
        \begin{itemize}
            \item \ref{plastick} - printer ran out of plastic before the print was finished;
            \item \ref{unstick} - part of the print unstuck from the printing base, but the rest of print remained undisturbed; 
            \item \ref{retraction} - speed of the retraction has been set too low (to 0.5);
            \item \ref{bowden} - during the printing, the Bowden tube fell out from its place;
            \item \ref{arm} - during the printing, the arm of printer head has been detached from magnets holding it in the place;
        \end{itemize}
    \item \textbf{variant (b)}:
        \begin{itemize}
            \item \ref{removal} - during the printing, part of the print has been removed.
        \end{itemize}
\end{enumerate}

\section{Data organization} \label{org}

All of the data is uploaded to the repository on GitHub\footnote{https://github.com/joanna-/3D-Printing-Data}. Structure of the directories is as follows:

\hfill \break

\begin{minipage}{0.5\textwidth}
  \dirtree{%
    .1 3D-Printing-Data.
    .2 four\_towers.
    .3 arm\_failure.
    .3 bowden.
    .3 plastic.
    .3 proper.
    .3 retraction.
    .3 unstick.
    .3 towers\_v01.stl.
    .2 four\_towers\_no\_base.
    .3 removing\_part.
    .3 proper.
    .3 towers\_only\_v01.stl.
}
\end{minipage}

\hfill \break

Each subdirectory contains zipped directory of the two files: csv data and json data. Apart from that, \textit{four\_towers} and \textit{four\_towers\_no\_base} contain \textbf{\.stl} files with the printing schemas.




\subsection{Custom measurements}

Custom measurements are stored in files with \textbf{csv} format without header line and with a standard comma separator. Each file contains the following columns:

\begin{itemize}
    \item \textbf{0} - data id assigned as the relative time set up on the device;
    \item \textbf{1-3} - 3-dimensional data from the printing base (X, Y, Z axis respectively);
    \item \textbf{4-6} - 3-dimensional data from the head accelerometer (X, Y, Z axis respectively);
    \item \textbf{7} - tension measurements;
    \item \textbf{8} - time stamps (in milliseconds).
\end{itemize}

\begin{lstlisting}[language=Python, caption=Csv data reading, frame=single, label=csv]
import pandas as pd
df = pd.read_csv('accel.txt', names=['data_id', 'accel0X', 'accel0Y', 'accel0Z', 'accel1X', 'accel1Y','accel1Z', 'tension', 'timestamp'])
df['time'] = pd.to_datetime(df['timestamp'], unit='ms')
df['tension'] = 0.650 * (df['tension'] - 2166)
\end{lstlisting}

In \textit{Listing \ref{csv}}, we present suggested way of reading the data using \texttt{pandas} library. The last line is needed as the effect of measurement device calibration - applying given formula scales the filament force data to start from 0. 

\subsection{Measurements from Duet Web Control}

Data that is provided by the web interface is in \textbf{json} file, in which each line contain different data sample object. The object includes many different key-values pairs and we will not discuss all of them in this paper as it is described in the device documentation. Below, we present several ones, that were useful during our analysis:

\begin{itemize}
    \item \textbf{coords.xyz} - array of length three with current \texttt{(x, y, z)} position of printing head;
    \item \textbf{status} - can be one of the following values:
    \begin{enumerate}
        \item \texttt{I} - for idle state;
        \item \texttt{P} - for printing phase;
        \item \texttt{T} - for temporary when printer is getting ready to printing phase;
    \end{enumerate}
    \item \textbf{temps.bed.current} - current temperature of the bed;
    \item \textbf{temps.extra} - among all contain MCU temperature;
    \item \textbf{currentLayer\textsuperscript{*}} - current printed layer - available only in some datasets.
\end{itemize}

In \textit{Listing \ref{json}}, we present suggested way of reading the json data using \texttt{pandas} library. Normalization of json data results in the flat (not nested) table structure for the obtained data. 

\begin{lstlisting}[language=Python,caption=Json data reading, label=json, frame=single]
import pandas as pd
from pandas.io.json import json_normalize
df = pd.read_json('interface.json', lines=True)
df = json_normalize(df.to_dict('records'))
\end{lstlisting}

\section{Exemplary Data analysis} \label{analysis}

\begin{table*}[ht]
\centering
\begin{tabular}{|l|c|l|}
\hline
\multicolumn{1}{|c|}{\textbf{failure type}} & \textbf{symptoms} & \multicolumn{1}{c|}{\textbf{brief explanation}} \\ \hline
\textit{finish of plastic} & \multirow{2}{*}{\begin{tabular}[c]{@{}c@{}}decrease of intrusion \\ power\end{tabular}} & there is no more plastic to intrude \\ \cline{1-1} \cline{3-3} 
\textit{Bowden tube fallout} &  & \begin{tabular}[c]{@{}l@{}}there is no friction with the print - plastic \\ doesn't reach printed model\end{tabular} \\ \hline
\textit{wrong retraction (0.5)} & \multirow{2}{*}{printing base jolting} & too much plastic hooks on the next layers \\ \cline{1-1} \cline{3-3} 
\textit{unsticking of the model} &  & printing head hooks on the rolled print \\ \hline
\textit{arm failure} & printing head angle change & detachment of arm causes head to tilt \\ \hline
\end{tabular}
\caption{Symptoms characteristic of the printing failures.}
\label{symptoms}
\end{table*}

Provoked failures cause different symptoms that can be detected with the data analysis. Different failures may have similar symptoms depending on their type and therefore inferring the initial cause can require more complex analysis. In this section we present very basics of analysis and show three types of symptoms related to five types of failures. The summary of failure-symptom correlation is presented in the \mbox{table \ref{symptoms}}.

\begin{figure*}
    \begin{center}
    \begin{subfigure}[t]{0.8\textwidth}
        \includegraphics[width=\textwidth]{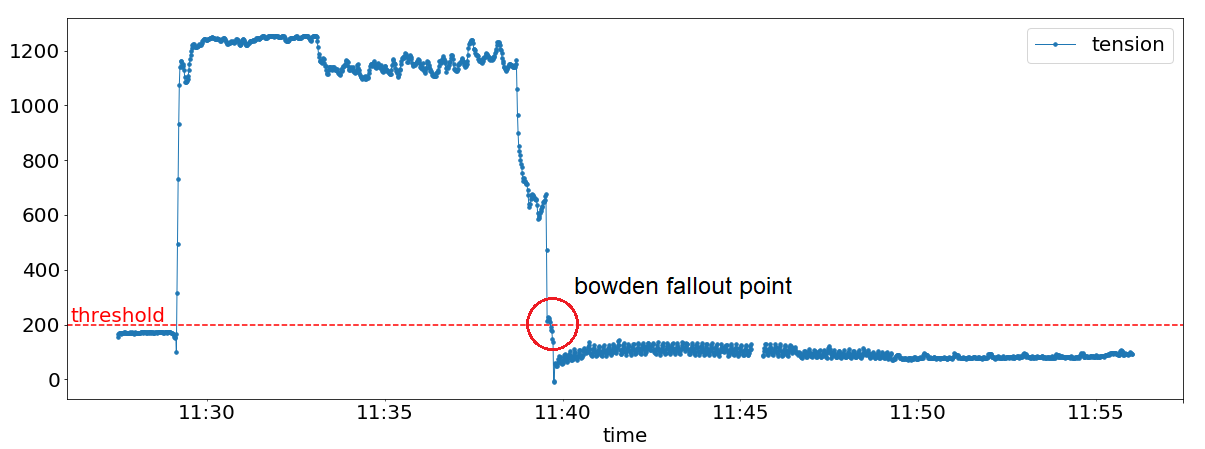}
        \caption{Tension values for the print with Bowden tube fallout.}
        \label{tensionplot}
    \end{subfigure}
    \end{center}
    \begin{center}
    \begin{subfigure}[t]{0.8\textwidth}
        \includegraphics[width=\textwidth]{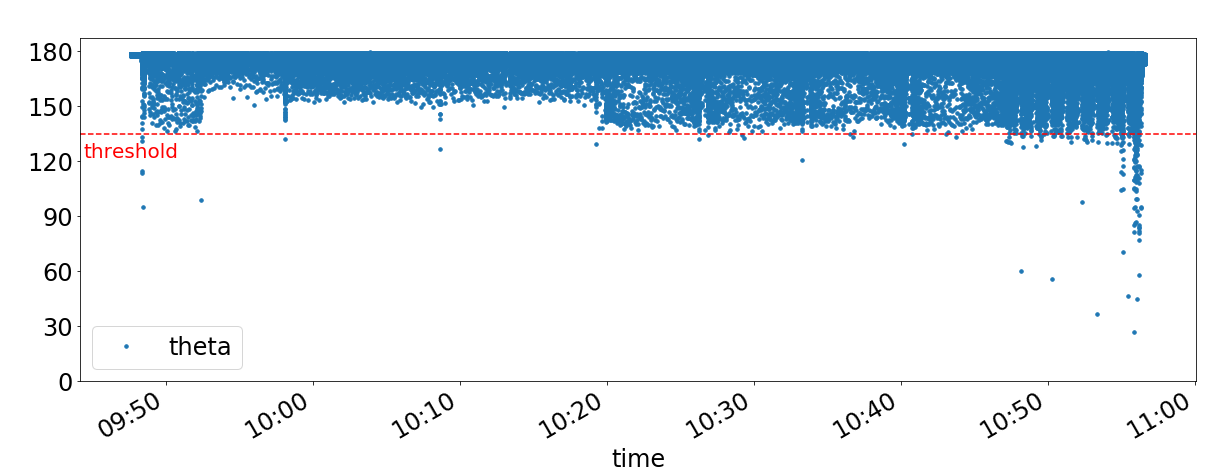}
        \caption{Tilt angle values for the print with head arm detachment.}
        \label{thetaplot}
    \end{subfigure}
    \end{center}
\caption{Plots presenting symptoms of printing anomalies}
\label{symptomsplots}
\end{figure*}

Figure \ref{symptomsplots} presents two different plots that show some of the aforementioned symptoms. Figure \ref{symptomsplots}a shows the situation where the filament feeding force dropped rapidly at time 11:40. That symptom may suggest that the filament is over or there was severe mechanical problem - in this case the Bowden tube fallout. Figure \ref{symptomsplots}b shows the tilt angle of the print head during printing. Values different from 180 degrees are caused by the fact that the angle is calculated on the basis of the accelerometer placed on the head, which is affected by the force of gravity and acceleration resulting from the movement of the head during printing. At 11:00 a significant change in the value of the graph can be observed on the chart indicating mechanical damage to the printer. In this case, the arm fixing the printing head in the delta system is damaged.

The presented analysis is only an example of using data to analyze the work of a 3D printer.

\section*{Summary}
The article presents the possibilities offered by the use of IoT devices in industry 4.0. Retrofitting machines with additional sensors and devices analyzing their work in real-time can provide valuable information about their work.

IoT devices such as those offered by \textit{FogDevices Platform} allow to simplify the process of adding sensors and analyzing data on the edge, near the sensors without sending them to the computational clouds.

The article presents data collected during the operation of the 3D printer, including typical errors. The collected data can be used to develop advanced algorithms for detection and prediction of failures.

\section*{Data usage}
The dataset is under Creative Commons Attribution 4.0 International license. Please cite this paper if you use it.

\section*{Acknowledgment}
The research presented in this paper was supported by the National Centre for Research and Development (NCBiR) under Grant No. LIDER/15/0144/L-7/15/NCBR/2016.

\bibliographystyle{IEEEtran}
\bibliography{bibliography}

\end{document}